\documentclass[prb,showpacs,twocolumn]{revtex4}
\usepackage{amsfonts}
\usepackage{amsmath}
\usepackage{amssymb}
\usepackage{graphicx}

\begin{document}

\title{Manipulation of a single magnetic atom using polarized single electron transport in a double quantum dot}
\author{Wenxi Lai and Wen Yang}
\affiliation{Beijing Computational Science Research Center, Beijing 100094, China}

\begin{abstract}
We consider theoretically a magnetic impurity spin driven by polarized electrons tunneling through a double quantum dot system. Spin blockade effect and spin conservation in the system make the magnetic impurity sufficiently interact with each transferring electron. As a results, a single collected electron carries information about spin change of the magnetic impurity. The scheme may develop all electrical manipulation of magnetic atoms by means of single electrons, which is significant for the implementation of scalable logical gates in information processing systems.
\end{abstract}

\pacs{76.30.Da, 72.25.Pn, 42.50.Dv, 73.23.Hk}
\maketitle

\begin{center}
\textbf{I. Introduction}
\end{center}

Magnetic atoms are critical spin systems which have potential applications in data storage and quantum information processing.~\cite{Thomas,Loth,Hanson:2008} In particular, using electrons to manipulate magnetic atoms is a natural step towards the implementation of scalable memory units for future integrated circuits. In dilute II-VI semiconductor quantum dots (QD), interaction between a manganese (Mn) atom and a carrier can be effectively described with the $sp-d$ exchange interaction.~\cite{Besombes,Reiter,Leger} Based on the impurity-carrier coupling, electrical control of the single magnetic atom is feasible by injecting different charges in the magnetic atom doped QD.~\cite{Leger}

As shown both in experiment~\cite{Loth} and theory,~\cite{Delgado} electrons can directly tunnel through a Mn atom by taking its spin states. The ground state and excited states of the Mn atom can be identified with the current, since they support different conductance. However, in the tunneling from a STM tip to $Cu_{2}N$ surface through an individual Mn atom, the Mn spin spontaneous relaxation is more frequent than excitation by tunneling electrons.~\cite{Loth} Compared with the case in high dimensional bulk material, the life time of the Mn spin is much longer in a QD.~\cite{Scheibner} Therefore, coherent electrical manipulation of the Mn spin is possible in low dimensional nanostructure. In a QD doped with a single Mn atom, charge and conductance of the single electron tunneling can be related to the spin state of the Mn atom.~\cite{Rossier} Whereas, it is hard to exactly connect a quantum state of the magnetic atom to a single electron in the above QD system, which should be a problem required to be solved for future quantum information processing.

In this paper, we propose a scheme for all electrical manipulation of the magnetic impurity spin, scaling the number of driving electrons down to one. To this end, we consider two inter-coupled semiconductor QDs, in which one QD contains a single magnetic impurity and couples to a spin polarized electron source. The other QD is localized in a homogeneous magnetic field and connected to a normal conductance, playing a role of a spin filter. In previous study,~\cite{Leger,Loth,Delgado,Rossier} electrons transport through the magnetic atom without any qualification to their spins, as a result, interaction between the magnetic atom and electrons are very weak. In contrast, the QD spin filter induces spin dependent tunneling, which makes sure each electron would be completely flipped by the magnetic atom before it passes through the filter. Therefore, each collected electron can be correlated with the change of spin state in the magnetic atom. There are several facts that are very beneficial for the realization of our scheme. First, coupled two QDs can be fabricated, doping a single magnetic ion in one of them.~\cite{Goryca} Second, spin life time of an electron in II-VI semiconductor quantum dots can be long enough. Electron relaxation time in a similar system was reported to be 50 $ns$ in a previous work.~\cite{Feng} In a QD imbedded a magnetic atom, a longer life time of around 1 $\mu s$ was predicted.~\cite{Petrovic} Third, Magnetic atom such as Mn impurity with relaxation time from 1 $\mu s$ to 0.4 $ms$ was observed in experiment.~\cite{Goryca,Le-Gall} The time is longer than driving time of the magnetic atom, which will be shown later in the present work. Forth, the technique for real time detection of single electron tunneling has been well developed recently.~\cite{Bylander,Lu,Gustavsson,Pekola} Fifth, polarized electron current is available from several sources, for instance, QD spin splitter under the premise of local magnetic field,~\cite{Recher,Hanson:2004} ferromagnetic leads,~\cite{Hyde,Dirnaichner} graphene or carbon atom wires.~\cite{Li,Karpan} Recently, nuclear spins of donor atoms such as phosphorus and $^{29}$Si have been coherently controlled and read using bounded electrons in these donors and the spin to charge conversion. Electromagnetic field was applied to initialize these nuclear spins.~\cite{Pla:2012,Pla:2013,Pla:2014} Comparing with these early works, the main advance in our present protocol is that the magnetic atom would be manipulated all electrically. In other words, a magnetic atom can be controlled only using a single electron transistor without the application of any electromagnetic field. Even the initialization of the impurity spin can be progressed using the single electron tunneling in principle.

\begin{figure}
\includegraphics[width=8.7cm]{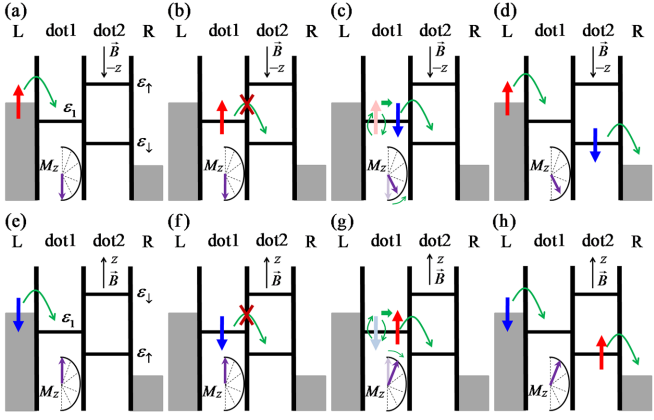}\\
\caption{(Color on line) Schematic illustration of the principle in our model. In (a) - (d) the applied magnetic field is in down direction and up polarized electrons are injected into the double QD. In (e)-(h), the external magnetic field is changed to be up direction and down polarized electrons are injected.}
\label{system}
\end{figure}

\begin{center}
\textbf{II. MODEL AND THEORY}
\end{center}

Our model is illustrated in Fig.~\ref{system}. The two QDs are denoted by dot 1 and dot 2 with ground orbital levels $ \varepsilon_{1}$ and $\varepsilon_{2}$, respectively. Both electron polarization in the left lead and the external magnetic field $\vec{B}$ that applied on the dot 2 are assumed to be parallel to the QD growth direction $z$. Electrons injected from the left lead into the dot 1 are coupled to the magnetic atom by the ferromagnetic Heisenberg type spin exchange interaction. We describe spin of the magnetic atom with mean value of the spin along $z$ direction $\langle \hat{M}_{z} \rangle$, where $\hat{M}$ is the magnetic atom spin operator and the bracket indicates average over quantum states of the system. Bias voltage and the magnetic field is tuned that just the lowest levels of the two QDs fall within the bias window $\mu_{L}>\varepsilon_{1}$, $\varepsilon_{\downarrow}$ (or $\varepsilon_{\uparrow}$)$>\mu_{R}$, where the indexes $\downarrow$, $\uparrow$ indicate electron states with spin up and down, respectively. In addition, the intra-dot Coulomb blockade energies $U_{1}$, $U_{2}$ corresponding to the dot 1 and the dot 2 are assumed to be much larger than other energy scales, which yields only single electron occupation is involved in either of the QDs. It remarkably simplifies our model and calculation.

In Figs.~\ref{system}(a)-(d), the magnetic field in this configuration is applied along the $-z$ direction with value $\vec{B}=(0, 0, -B)$. It means, in dot 2, the ground state level is $\varepsilon_{\downarrow}=\varepsilon_{2}-g^{*}\mu_{B}B/2$ and the first excited level is  $\varepsilon_{\uparrow}=\varepsilon_{2}+g^{*}\mu_{B}B/2$, where $g^{*}$ is the Lande $g$-factor of electron in the QD, $\mu_{B}$ is the Bohr magneton. The system requires spin up electrons that injected from the left lead into the dot 1. We take energy levels that satisfy $\varepsilon_{\uparrow} > \mu_{L}$ and $\varepsilon_{\uparrow}-\varepsilon_{1} \gg \hbar\Omega$, where $\hbar\Omega$ is inter-dot coupling strength. This energy structure forms a spin conditioned repulsive potential which ensures that the spin up electron is forbidden to enter the dot 2 until its spin is flipped to be upside down due to its coupling to the magnetic atom. As soon as the electron spin changes to be down, it would be allowed to pass through the dot 2 and collected in the right lead. At the same time, the spin of magnetic atom would change from $\langle \hat{M}_{z} \rangle$ to $\langle \hat{M}_{z} \rangle+\hbar$. If a spin down electron is injected from the left lead, it directly transports through the double dots without any change in the spin of the magnetic atom. However, if the left lead is a fully up polarized electron source, each passed electron would be connected to the spin change of the magnetic atom. In this case, due to spin conservation, if spin of the magnetic atom is changed from $\langle \hat{M}_{z} \rangle$ to $\langle \hat{M}_{z} \rangle+n_{\downarrow}\hbar$, then the number of electrons detected in the right lead should equal to $n_{\downarrow}$ and their spins are down polarized.

The magnetic atom can also be driven reversely, which is presented in Figs.~\ref{system}(e)-(h). Since the magnetic field is turned to be in $z$ direction here, the spin filter only allows spin up electrons tunnel through the dot 2. The input electrons are required to be in spin down state and output electrons are expected to be in spin up state. Each transported electron contributes to the spin of the magnetic atom with the value $-\hbar$. It yields the impurity magnetization orient in the reversal way, which means spin of the magnetic atom would be changed from $\langle \hat{M}_{z} \rangle$ to $\langle \hat{M}_{z} \rangle-n_{\uparrow}\hbar$ with collection of $n_{\uparrow}$ spin up electrons.

To give a quantitative description to the model, we use the following Hamiltonian~\cite{Reiter,Becker}

\begin{eqnarray}
\hat{H}=\hat{H}_{d1}+\hat{H}_{d2}+\hat{H}_{d12}+\hat{H}_{lead}+\hat{H}_{tun},
\label{eq:Hamiltonian}\end{eqnarray}
where the Hamiltonian for dot 1 is
\begin{eqnarray}
\hat{H}_{d1}=\varepsilon_{1}\hat{n}_{1}+U_{1}\hat{n}_{1\uparrow}\hat{n}_{1\downarrow}-j_{e}\vec{\hat{M}} \cdot \vec{\hat{S}}_{1},
\end{eqnarray}
and the Hamiltonian for dot 2 is
\begin{eqnarray}
\hat{H}_{d2}=\varepsilon_{2}\hat{n}_{2}+U_{2}\hat{n}_{2\uparrow}\hat{n}_{2\downarrow}+g^{*}\mu_{B}\vec{B} \cdot \vec{\hat{S}}_{2}.
\end{eqnarray}
The inter-dot tunneling Hamiltonian reads
\begin{eqnarray}
\hat{H}_{d12}=\hbar \Omega (\hat{n}_{12}+\hat{n}_{12}^{\dag}) + U \hat{n}_{1}\hat{n}_{2}.
\end{eqnarray}
Here, $\hat{n}_{i}=\hat{n}_{i\uparrow}+\hat{n}_{i\downarrow}$,  $\hat{n}_{i\uparrow}=\hat{c}^{\dag}_{i\uparrow}\hat{c}_{i\uparrow}$,  $\hat{n}_{i\downarrow}=\hat{c}^{\dag}_{i\downarrow}\hat{c}_{i\downarrow}$, $\hat{n}_{12}=\hat{n}_{12\uparrow}+\hat{n}_{12\downarrow}$, $\hat{n}_{12\uparrow}=\hat{c}^{\dag}_{1\uparrow}\hat{c}_{2\uparrow}$, $\hat{n}_{12\downarrow}=\hat{c}^{\dag}_{1\downarrow}\hat{c}_{2\downarrow}$. $\hat{c}_{i\uparrow}$ ($\hat{c}_{i\downarrow}$) is the annihilation operator of spin up (down) electron in dot $i$ $(i=1,2)$. $S_{1}$ and $S_{2}$ represent electron spin in the dot 1 and the dot 2, respectively. $U$ is the inter-dot Coulomb potential. The exchange coupling strength between the electron in dot 1 and the magnetic atom is given by $j_{e}=J|\psi_{0}(r_{M})|^{2}$ with exchange integral $J$ and the electron ground state wave function $\psi_{0}$ at the magnetic impurity position $r_{M}$.

The left and right electronic leads are described by the free electron baths with the Hamiltonian
\begin{eqnarray}
\hat{H}_{lead}=\sum_{k,\sigma; \alpha=L,R} \epsilon_{\alpha k} \hat{c}^{\dag}_{\alpha k \sigma} \hat{c}_{\alpha k \sigma}.
\end{eqnarray}
The dot 1 is coupled to the left leads and the dot 2 is coupled to the right lead by
\begin{eqnarray}
\hat{H}_{tun}= \sum_{k,\sigma}V_{L} \hat{c}^{\dag}_{L k \sigma}\hat{c}_{1 \sigma}+ \sum_{k,\sigma}V_{R} \hat{c}^{\dag}_{R k \sigma}\hat{c}_{2 \sigma}+H.c.,
\end{eqnarray}
with the left and right tunneling amplitudes $V_{L}$ and $V_{R}$, respectively.

Time evolution of electron transport through the double QD system is described by a quantum master equation which is derived based on the Hamiltonian~\eqref{eq:Hamiltonian} and the Liouville-von Neumann equation in the Born-Markov approximation. Since the transport in our system is a process of single electron sequential tunneling through QDs and works in the weak tunneling regime, the master equation is an effective approach to describe our model.~\cite{Gurvitz,Novotny,Braig} Equation of motion is given in terms of the reduced density matrix $\hat{\rho}$ of the system,

\begin{eqnarray}
\frac{\partial}{\partial t}\hat{\rho} &=&\frac{1}{i \hbar} [\hat{H}_{d1}+\hat{H}_{d2}+\hat{H}_{d12},\hat{\rho}]+\hat{\mathcal{L}}_{L}\hat{\rho}+\hat{\mathcal{L}}_{R}\hat{\rho}.
\end{eqnarray}

The Liouville super-operators, $\hat{\mathcal{L}}_{L}$ and $\hat{\mathcal{L}}_{R}$, acting on the density matrix $\hat{\rho}$ describe tunneling on the left and right side of the double dots, respectively. They are written as
\begin{widetext}
\begin{eqnarray}
\hat{\mathcal{L}}_{L}\hat{\rho}=\frac{1}{2}\sum_{\sigma}\Gamma_{L}^{\sigma} (\hat{f}_{L,\sigma}(\hat{c}^{\dag}_{1\sigma}\hat{\rho} \hat{c}_{1\sigma}-\hat{c}_{1\sigma}\hat{c}^{\dag}_{1\sigma}\hat{\rho} )+(1-\hat{f}_{L,\sigma})(\hat{c}_{1\sigma}\hat{\rho} \hat{c}^{\dag}_{1\sigma} -c^{\dag}_{1\sigma}c_{1\sigma}\hat{\rho} )+H.c.),
\end{eqnarray}
and
\begin{eqnarray}
\hat{\mathcal{L}}_{R}\hat{\rho}=\frac{1}{2}\sum_{\sigma}\Gamma_{R}^{\sigma} (\hat{f}_{R,\sigma}(\hat{c}^{\dag}_{2\sigma}\hat{\rho} \hat{c}_{2\sigma}-\hat{c}_{2\sigma}\hat{c}^{\dag}_{2\sigma}\hat{\rho} )+(1-\hat{f}_{R,\sigma})(\hat{c}_{2\sigma}\hat{\rho} \hat{c}^{\dag}_{2\sigma} -\hat{c}^{\dag}_{2\sigma}\hat{c}_{2\sigma}\hat{\rho} )+H.c.),
\end{eqnarray}
\end{widetext}
where the Fermi distribution function in the left lead is $\hat{f}_{L,\sigma}=(exp[(\varepsilon_{1}+U_{1}\hat{n}_{1 \bar{\sigma}}-\mu_{L})/k_{B} T]+1)^{-1}$ and in the right lead is  $\hat{f}_{R,\sigma,}=(exp[(\varepsilon_{\sigma}+U_{2}\hat{n}_{2 \bar{\sigma}}-\mu_{R})/k_{B} T]+1)^{-1}$, depending on the charging energy $U_{i}$ conditioned by the occupation, $n_{i \bar{\sigma}}=c_{1 \bar{\sigma}}^{\dag }c_{1 \bar{\sigma} }$, of electron with spin $\bar{\sigma}$. Here, the spin index $\bar{\sigma}$ is defined that if $\sigma=\uparrow (\downarrow)$, then $\bar{\sigma}=\downarrow(\uparrow)$. The tunneling rates are spin dependent given by $\Gamma_{\alpha}^{\uparrow}=\Gamma_{\alpha}(1+P_{\alpha})/2$ and  $\Gamma_{\alpha}^{\downarrow}=\Gamma_{\alpha}(1-P_{\alpha})/2$ with the current polarization  $P_{\alpha}=(I_{\alpha}^{\uparrow}-I_{\alpha}^{\downarrow})/(I_{\alpha}^{\uparrow}+I_{\alpha}^{\downarrow})$. Here, $I_{\alpha}^{\sigma}$ is the current with pure spin $\sigma$ on the side of $\alpha$ ($\alpha=L, R$). The bare tunneling rates can be expressed as $\Gamma _{\alpha }=2 \pi \hbar \left\vert t_{\alpha }\right\vert ^{2}N_{\alpha }(\epsilon )$ with the density of states $N_{\alpha }(\epsilon )$ of electrons at energy $\epsilon$.

\begin{figure}
\includegraphics[width=9cm]{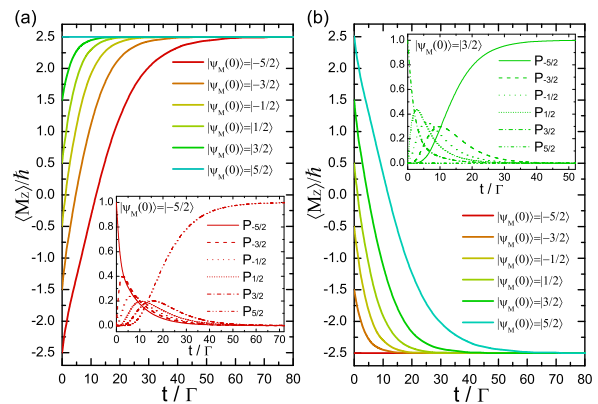}\\
\caption{(Color on line) Time evolution of the single impurity magnetization as a function of time. (a) The applied external magnetic field is $\vec{B}=(0,0,-B)$. (b) The applied external field is $\vec{B}=(0,0,B)$. The Parameters are $\mu_{L}=75\Gamma$, $\mu_{R}=-75\Gamma$, $\varepsilon_{1}=0$, $\varepsilon_{2}=62.5\Gamma$, $g^{*}\mu_{B}B=135\Gamma$, $j_{e}=3\Gamma$, $k_{B}T=12.5\Gamma$, $\Gamma_{L}=\Gamma_{R}=\Gamma$, $P_{R}=0$, $\hbar\Omega=5\Gamma$, $U=10\Gamma$.}
\label{mpt6}
\end{figure}

\begin{figure}
\includegraphics[width=8cm]{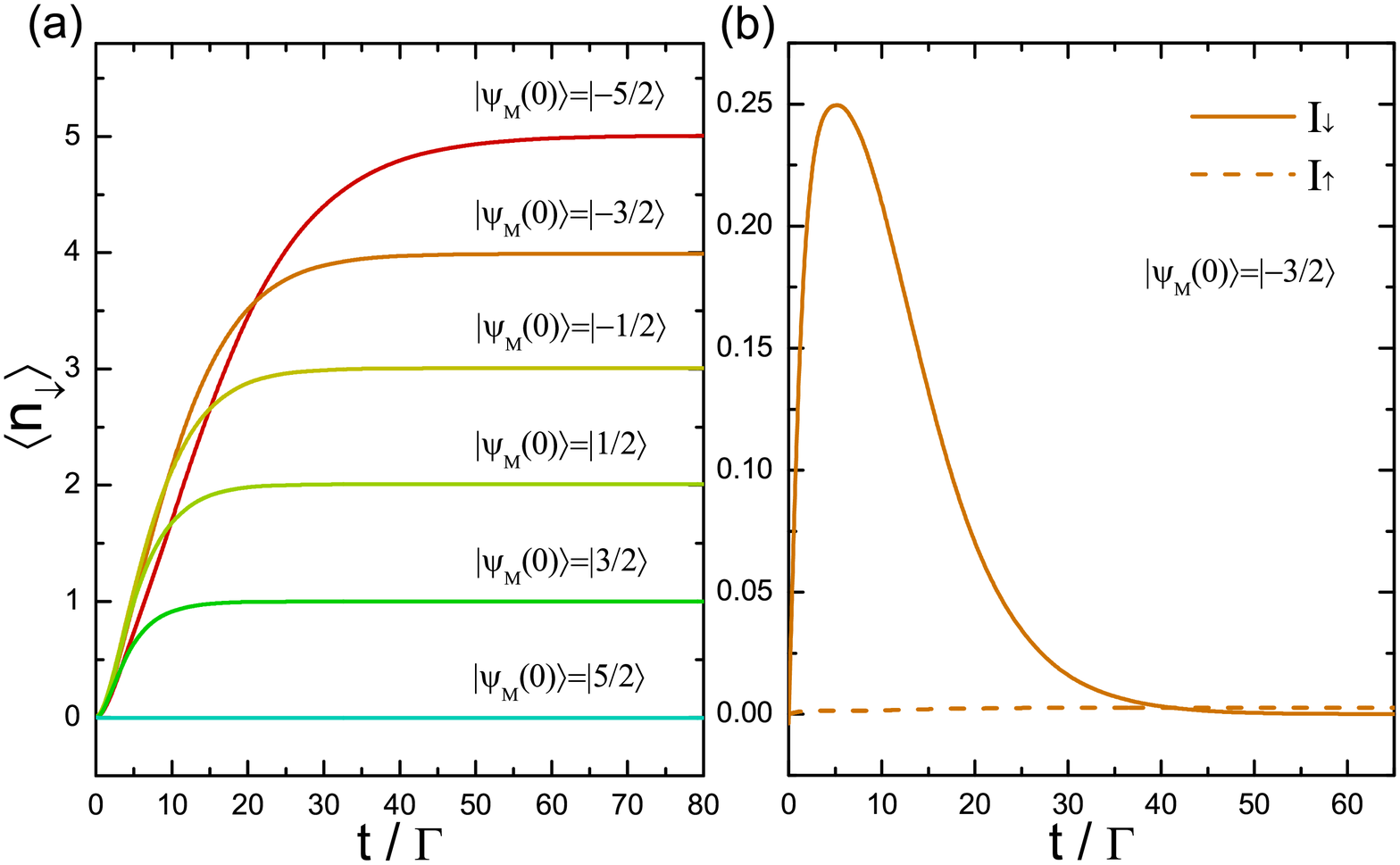}\\
\caption{(Color on line) (a) The number of spin down electrons collected in the right side of the system. (b) Current for spin up and down electron s as a function of time. The parameters are the same as that in Fig.~\ref{mpt6}}
\label{nit6}
\end{figure}

\begin{center}
\textbf{III. MANIPULATION OF A MAGNETIC ATOM WITH SPIN M=5/2}
\end{center}

The Hilbert space of the double QD system is generated by the basic vectors $|i\rangle_{1} |m\rangle_{1} |j\rangle_{2}$, where $|i\rangle_{1}$ and $|j\rangle_{2}$ represent electronic state in dot 1 and dot 2, respectively. Here, $i,j=0$ denote the empty state, $i,j=\downarrow, \uparrow$ indicate occupation states of single electron with spin down and spin up, respectively. $ |m\rangle_{1}$ is eigenstate of the impurity spin operator $\hat{M}_{z}$ with eigenvalues $m=-M, -m+1, ... , M$.

First, we consider a typical single magnetic atom which displays a spin of $M=5/2$ in the QD, such as Fe or Mn. These magnetic atoms have six quantized spin states $|m\rangle$, with corresponding eigenvalues $m=-5/2$,$-3/2$,$-1/2$,$1/2$,$3/2$,$5/2$.~\cite{Besombes} In Fig.\ref{mpt6}(a), initial state $|\psi_{M}(0)\rangle$ of the magnetic atom is set to be any of the six quantized states. In all cases the magnetic atom is driven to the final state $|5/2 \rangle$. In average value of the Mn spin $\langle M_{z} \rangle = Tr[\hat{M}_{z}\hat{\rho}]$ trace is taken over the basic vectors $|i\rangle_{1} |m\rangle_{1} |j\rangle_{2}$. When the impurity spin changes to $|5/2\rangle$, it becomes parallel to the injected electron spin and there will be no spin flip in the later time. Then the electron tunneling should be switched off and the spin down polarized current decreases to be zero as plotted in Fig.~\ref{nit6} (b). The figures also show that orientation time of the Mn spin is a few tens of $ns$ when one takes a tunneling characteristic time of $\Gamma^{-1} \sim 1 ns$. It is comparable to the time scale of a previously reported optical control.~\cite{Le-Gall} Since each electron contributes to the spin change of the magnetic atom with momentum $\hbar$, the magnetic atom initialized in the states $|\psi_{M}(0)\rangle=|-5/2 \rangle$, $|-3/2 \rangle$, $|-1/2 \rangle$, $|1/2 \rangle$, $|3/2 \rangle$, $|5/2 \rangle$ leads to finite electrons collected in the right lead with definite numbers $\langle n_{\downarrow} \rangle=5,4,3,2,1,0$, respectively. It can be seen in Fig.~\ref{nit6}(a) for corresponding initial spin states. The collected electron number is calculated using the formula

\begin{eqnarray}
\langle n_{\sigma} \rangle=\frac{1}{e}\int_{0}^{\infty}I_{\sigma}(t)dt,
\end{eqnarray}
where $I_{\sigma}(t)$ is current on the right side of dot 2. The current is derived from the charge fluctuating in the two dots $ d(\langle n_{1} \rangle+\langle n_{2} \rangle)/dt=\sum_{\sigma=\uparrow,\downarrow}(I^{\sigma}_{L}-I^{\sigma}_{R})/e$ with the replacement $I_{\sigma}(t)=I^{\sigma}_{R}$. The reversal time evolution of the magnetization is shown in Fig.~\ref{mpt6} (b).

\begin{figure}
\includegraphics[width=8cm]{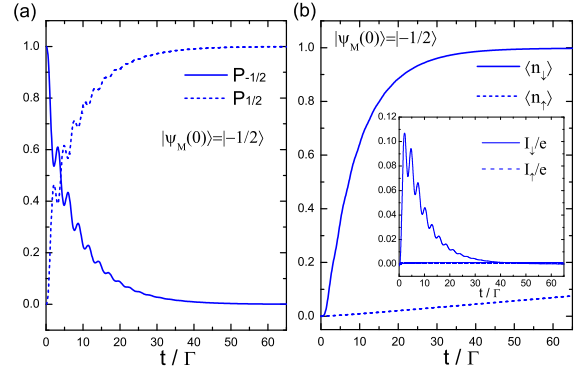}\\
\caption{(Color on line) (a) Probabilities of the magnetic atom in its spin states $|-1/2\rangle$ and $|1/2\rangle$. The magnetic atom is initially in the state $|-1/2\rangle$. (b) The number of collected electrons with spin up and spin down corresponding to the case (a). The parameters are $\mu_{L}=70\Gamma$, $\mu_{R}=-70\Gamma$, $\varepsilon_{1}=0$, $\varepsilon_{2}=57.5\Gamma$, $g^{*}\mu_{B}B=125\Gamma$, $j_{e}=2\Gamma$, $k_{B}T=10\Gamma$, $\Gamma_{L}=\Gamma_{R}=\Gamma$, $P_{L}=1$, $P_{R}=0$, $\hbar\Omega=3\Gamma$, $U=10\Gamma$.}
\label{pnit2}
\end{figure}

\begin{center}
\textbf{IV. MANIPULATION OF A MAGNETIC ATOM WITH SPIN M=1/2}
\end{center}

Now, we consider a magnetic ion with spin $M=1/2$, such as Cu$^{2+}$. This kind of impurity has particular meaning that its maximum magnetization difference is $\hbar$, from $-\hbar/2$ to $\hbar/2$ or by inverse. In this case, only a single electron is involved in the tunneling. As shown in Fig.\ref{pnit2}(a), when we set initial state of the magnetic atom to be $|-1/2\rangle$, it is transformed into $|1/2\rangle$. At the same time, the up polarized current is changed to be down polarized current after interacting with the magnetic atom. The down polarized current is expected to be collected in the right side, which is characterized by a singe electron with spin down as shown in Fig.\ref{pnit2}(b). To show the collected charge is really limited, average current as a function of time is plotted in the inset of Fig.\ref{pnit2}(b). The down polarized current sharply increases and then disappears slowly with a small fluctuation. The up polarized current is negligible weak. From the point of application, it is a very good system for date storage, where only single electron of current and the two states of the magnetic atom is correlated.

The small fluctuation in the current is occurred due to the back action from the system. Since charge transfer through the double dots requires electron spin flip, the amplitude of current is proportional to rate of electron spin change. The rate of spin change is determined by the strength of spin coupling between the magnetic atom and electrons. Therefore, we can deduce that the oscillations observed in Fig.\ref{pnit2} is contributed from the coherent coupling between the magnetic atom and an individual electron. Indeed, character of the small oscillation can be tuned by changing the spin coupling strength $j_{e}$. Actually, the back action effect can also be seen in the former situation for $M=5/2$. However, considering six spin states of the magnetic atom are involved in the exchange interaction and the larger couplings $j_{e}$ and $\Omega$ are taken, the small oscillation during the evolution of the spin states and current is too smooth to be observed.

\begin{figure}
\includegraphics[width=10cm]{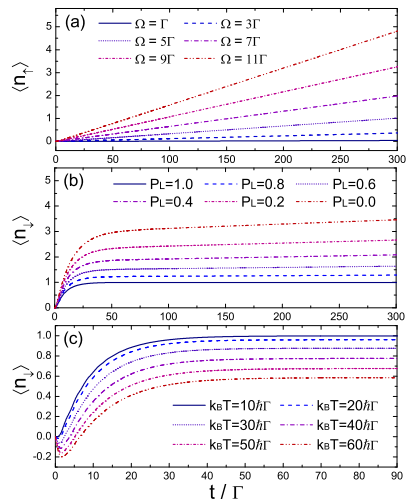}\\
\caption{(Color on line) (a) Number of spin up electrons collected in the right side of the double QD with different inter-dot coupling strength, $k_{B}T=10\Gamma$, $P_{L}=1$. (b) Number of spin down electrons for different spin polarization in the left lead, $k_{B}T=10\Gamma$, $\hbar\Omega=3\Gamma$. (c) Number of collected spin down electrons at different temperature, $\hbar\Omega=3\Gamma$, $P_{L}=1$. The rest parameters are the same as that in Fig.~\ref{pnit2}.}
\label{noptt2}
\end{figure}

The above results are described in a relatively ideal situation. At the end of this section, let us talk about some more practical cases. Fig.~\ref{noptt2}(a) reveals that increase of the inter-dot tunneling strength excites a spin up electron from the dot 1 into the excited level $\varepsilon_{\uparrow}$ of the dot 2. It leads to the unexpected spin up electron leakage from the double dot system into the right lead. To restrain the electron leakage through the excited level $\varepsilon_{\uparrow}$, we suggest that one can take a relatively small inter-dot coupling $\hbar\Omega$ which is required to be much smaller than the decoupling $\varepsilon_{\uparrow}-\varepsilon_{1}$.

The system is also sensitive to polarization of the right lead. In Fig.~\ref{noptt2}(b), it is illustrated that when the polarization is not pure, the number of electrons collected in the right lead is larger than the expected value. Since non-purely polarized current contains electrons with different spin orientations, thereby some electrons could not be blocked by dot 2.

The negative number of electrons in Fig.~\ref{noptt2}(c) implies that electrons in the right lead have certain probability to flow into the QDs at the beginning of tunneling due to the thermal excitation, since the QDs is empty initially. After the system reaches steady state, the higher the temperature, the larger electron distribution would be in the QDs. The electron staying in the QDs comes from the leads. Therefore, net number of electrons collected in the right lead decreases when temperature increases.

\begin{center}
\textbf{V. DISCUSSIONS}
\end{center}

In experiment, to realize electron transport through a magnetic atom doped II-VI semiconductor QD is still a challenge. There are some well developed experimental backgrounds, for instance, the spin dependent electron transport through a quantum well that containing dilute magnetic material,~\cite{Slobodskyy} and resonant tunneling through a CdSe self-assembled QD with Mn ions.~\cite{Gould} In addition, an attempt to orient a Mn spin using the charge that transported from a neighboring QD has been successfully realized.~\cite{Goryca}

To create a local magnetic field is a key technical problem for the experimental implementation of our proposal. The left lead is a polarized electron source, so an external magnetic field may be required on it. In addition, the right QD needs a magnetic field to create a Zeeman splitting for an electron. At the same time, these magnetic fields must have negligible affect to the left QD and the right lead. Therefore, some local magnetic fields are necessary in the nanostructure. To this end, a proper quantity of external magnetic field can be exerted on the left lead for the generation of spin polarized electron source. As the QD is a zero dimensional nanoscale system, our suggestion is using a magnetic grain, such as Co grain,~\cite{Nogueira,Dzemiantsova} to create magnetic field on the right QD. To explore its effectiveness we estimate the field strength of a typical Co grain which consists of hundreds of atoms. We assume the Co grain is a uniformly magnetized sphere. Then, the field of the grain can be obtain from the formula $B=\frac{2}{3}\mu_{0} \mathbb{M}$, $\mu_{0}$ is permeability of free space and $\mathbb{M}$ is magnetization of the grain.~\cite{Griffiths} We take that average magnetic moment per Co atom is $1.5 \mu_{B}$. In real materials the local magnetic moment is always larger than this value.~\cite{Nogueira} We take diameter of a Co atom is about $0.15 nm$. Then, magnetic field of the Co grain is estimated to be not less than $B=6.6 T$. By increasing size of the Co grain, more strong field can be obtained. In a CdTe semiconductor QD, the corresponding Zeeman splitting of an electron reaches $|g^{*}\mu_{B}B|=0.64 meV$, where the $g$-factor is given by $g^{*}=-1.67$ in this material.~\cite{Qu:2005} To guarantee the left QD and the right lead is not effectively influenced by the surrounding magnetic field, magnetic shielding may be applied here to protect the spreading field. One kind of the magnetic shielding materials is superconducting chip which expels magnetic field via the Meissner effect. Another kind shielding material is certain high permeability metal alloy which, in contrast the superconductor, draws the field into themselves. Taking an example with the high permeability shielding with permeability $\mu$, field in the shielded volume is $\frac{9 b^{3} \mu_{0}}{2 (b^{3}-a^{3}) \mu}$ times smaller than the outside field.~\cite{Jackson} For the convenience of quantitative estimation, the shielding material here is assumed to be a spherical shell with inner radius $a$ and outer radius $b$. When the permeability $\mu$ is large enough, good shielding of the field in the shielded area can be achieved. In practice, as the QD couples to an electronic reservoir and another dot, the shielding material may be not absolutely closed with a lower efficiency.

In the Hamiltonian of our model, spin exchange interaction between the two QDs is not considered. In fact, when each quantum dot contains one electron, the exchange interaction does not play important role for the whole system. The reason is that the electron in dot 2 has definite spin direction. It has to gain large energy to change spin. However, it is hard for the electron in dot 1 to provide the large energy. Besides, the system is robust against the double occupancy in dot 1. Since current injected from the left lead is assumed to be pure spin polarized, two electrons with different spin states in dot 1 does not break the spin conservation as soon as dot 2 guarantees to output electrons with pure spin. Whereas, double occupancy in dot 2 induces spin leakage and, as a result, the correlation between electrons and the magnetic atom becomes weak. Even though, considering the high Coulomb blockade effect in either of the QDs, double occupancy in any dot is negligible small in our model.

As an information processing system, its characteristic times are very significant. There are two kinds of critical times, one of them is manipulation time $\tau_{s}$ of the system, another is lowest life-time bound $\tau_{i}$ of these information carriers, such as the magnetic atom and electrons. The manipulation time of the system indicates a time range during which an electron is emitted from the left lead and then collected in the right lead, at the same time control of the magnetic atom is accomplished. It is clear that spin life times of the information carriers are required to be, at least, longer than the manipulation time of the system, i.e. $\tau_{i}>\tau_{s}$. As mentioned in the introduction, a single Mn spin relaxation time from 1 $\mu s$ to 0.4 $ms$ in CdTe QD~\cite{Goryca,Le-Gall}, and electron life time from 50 $ns$ to 1 $\mu s$ in II-VI semiconductor systems are reported. Even in a QD including certain charges, the Mn atom relaxes in a time scale of about 100 $ns$.~\cite{Le-Gall:2010} As shown in section III and IV, manipulation time of the system is about 50 $ns$ for a characteristic time of the electron tunneling, $\Gamma^{-1}\sim 1 ns$. Furthermore, there is still a space for reducing the manipulation time of the system. Actually, properly increasing the dot-reservoir coupling (electron tunneling rate) or inter-dot coupling would improve the rate of control process. In this case, a more lower bound of the spin life time is allowed.

\begin{center}
\textbf{VI. CONCLUSIONS}
\end{center}

In the system, change of magnetic impurity spin is correlated to spin state and number of single electrons that tunneling through the two QDs. Based on this principle, we give the following predictions: (i) Our model works as an electron source in which number of emitted electrons can be determined beforehand by setting an initial state of the magnetic atom or using a magnetic atom with certain spin $M$. In particular, a single electron emitter is available using a magnetic atom with spin $1/2$. (ii) The number of polarized transferring electrons can be recorded in the spin state of the magnetic atoms. (iii) The change in spin state of the magnetic atom should be detected by counting the number of electrons that emitted from the double QDs. (iv) Spin state of the magnetic impurity can be controlled, in principle, by injecting suitable number of spin polarized electrons, and this controlling is reversal.

\begin{acknowledgments}
We appreciate the valuable discussion with Shuhui Zhang about the numerical calculation.
\end{acknowledgments}

\end{document}